\documentclass[referee]{aa} 
\usepackage{graphicx}
\usepackage{txfonts}
\begin{document}
\title{The exoplanet-host star iota Horologii: an evaporated member of the primordial Hyades cluster. }
\author{S. Vauclair\inst{1} 
\and M. Laymand \inst{1}
\and F. Bouchy \inst{2} 
\and G. Vauclair \inst{1}
\and A. Hui Bon Hoa \inst{1} 
\and S. Charpinet \inst{1} 
\and  M. Bazot \inst{3}}
\offprints{Sylvie Vauclair}
\institute{Laboratoire d'Astrophysique de Toulouse-Tarbes, CNRS, 
Universit\'e de Toulouse, 14 av. Ed. Belin, 31400 Toulouse, France
\and Institut d'Astrophysique de Paris, 75014 Paris, France
\and Centro de Astrophysica da Universidade do Porto, Porto, Portugal}
\titlerunning{ }

   \date{Received ; accepted }

 
  \abstract
   {}
   { We show that the exoplanet-host star iota Horologii, alias HD17051, which belongs to the so-called Hyades stream, was formed within the primordial Hyades stellar cluster and has evaporated towards its present location, 40 pc away.}
   { This result has been obtained unambiguously by studying the acoustic oscillations of this star, using the HARPS spectrometer in La Silla Observatory (ESO, Chili).}
   { Besides the fact that $\iota$ Hor belongs to the Hyades stream, we give evidence that it has the same metallicity, helium abundance, and age as the other stars of the Hyades cluster. They were formed together, at the same time, in the same primordial cloud.}
   {This result has strong implications for theories of stellar formation. It also indicates that the observed overmetallicity of this exoplanet-host star, about twice that of the Sun, is original and not caused by planet accretion during the formation of the planetary system.}

   \keywords{exoplanets - asteroseismology - stellar clusters - stars:abundances - stars: solar-type stars}

   \maketitle

\section{Introduction}

The exoplanet-host star iota Horologii, alias HD17051 or $\iota$ Hor, is located in the south hemisphere (Table 1) and belongs to the ``Hyades stream". Although situated about 40pc away from the Hyades cluster, this star moves inside the Galaxy in a direction similar to that of the cluster (Chereul et al. \cite{chereul99}, Grenon \cite{grenon00}, Chereul \& Grenon \cite{chereul01}, Montes et al. \cite{montes01}). 

The Hyades stream, which consists of a large number of stars with average galactic velocity components U~$\simeq$~-37 km.s$^{-1}$ and V~$\simeq$~-17 km.s$^{-1}$ has long been a subject of debate (U is the velocity towards the Galactic centre, V the velocity in the direction of the Galactic rotation, both with respect to the Sun, cf Famaey et al. \cite{famaey07}). From statistical considerations, Famaey et al (\cite{famaey07}) conclude that a large majority of these stars (about 85$\%$ of the stream for low-mass stars) are field-like stars sharing the Hyades galactic velocities because of galactic dynamical effects, whereas the remaining proportion (about 15$\%$) would have been evaporated from the primordial Hyades cluster. The exoplanet-host star $\iota$ Hor, with galactic velocity components U~$\simeq$~-32 km.s$^{-1}$ and V~$\simeq$~-17 km.s$^{-1}$ (Montes et al. \cite{montes01}, Nordstroem et al. \cite{nordstroem04}), could either be a field-like star captured by the stream or a Hyades evaporated star.

In the latter case, $\iota$ Hor should obviously have the same age, metallicity, and helium abundance as the Hyades stars. Asteroseismology, which consists in analysing the periodic acoustic motions of stellar surfaces and using them to probe the stellar interiors, was proposed as a unique test to provide an answer to this question (Laymand \& Vauclair \cite{laymand07}). Here we claim that the asteroseismic studies of this star indeed prove that it is an evaporated member of the primordial Hyades cluster. 

\section{Observations}

We observed the solar-type oscillations of $\iota$ Hor in November 2006 with the HARPS spectrometer (Mayor et al. \cite{mayor03}) at La Silla Observatory, Chili (Fig. 1). The oscillation frequencies were obtained using Fourier transform analysis of the radial velocity time series (Fig. 2). Up to 25 oscillation modes could be identified using the usual development in spherical harmonics. 

\begin{figure}
\includegraphics[width=9cm]{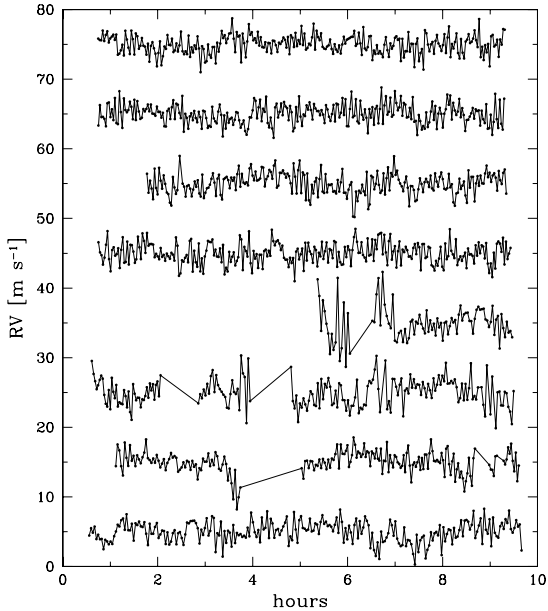}
\caption{Acoustic oscillations of the star $\iota$ Hor. The radial velocity curves obtained during the eight nights are displayed on this graph. 
For clarity they are arbitrarily shifted upwards by 10 m.s$^{-1}$ for each night. 
While the first and the four last nights are complete, the second, third, and fourth ones were partially cloudy, so only parts of them could be exploited. 
In this figure, the radial velocities have been corrected for long term variability
due to stellar activity. The 6-mn oscillations can be seen in the time series.}
\label{fig1}
\end{figure}

\begin{figure}
\includegraphics[width=9cm]{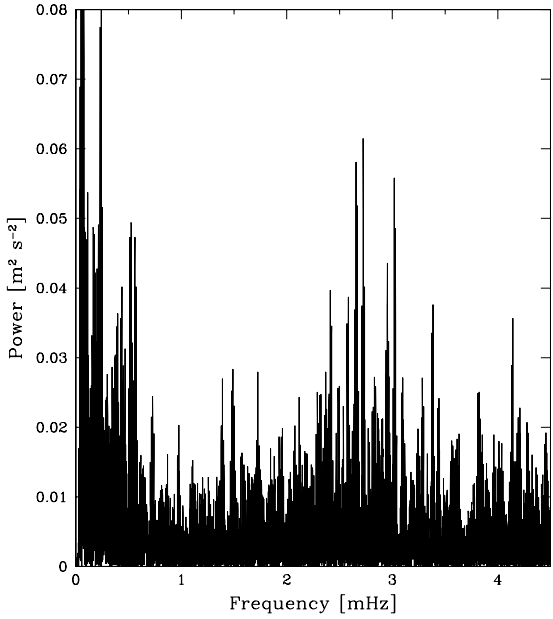}
\caption{ Frequency spectrum of the star $\iota$ Hor. This spectrum is the Fourier transform of the radial velocity curves. The peaks at low frequencies are signatures of the stellar activity, while those above 2 mHz correspond to the acoustic frequencies.}
\label{fig2}
\end{figure}

\begin{table*}
\caption{iota Horologii: observational data}
\label{tab1}
\begin{flushleft}
\begin{tabular}{cc} \hline
\hline
 right ascension & 02h 42mn 33s   \cr 
 declination & -50d 48' 01"\cr
 parallax (Hipparcos) & 58.00 $\pm$ 0.55 mas \cr
 luminosity (log L/L$_{\odot}$) & 0.219 $\pm$ 0.024 \cr
 effective temperature & 6136~K $\pm$ 34~K (1) \cr
    &  6252~K $\pm$ 53~K (2) \cr
    & 6097~K $\pm$ 44~K (3)\cr
 surface gravity (log g) & 4.47 $\pm$ 0.05 (1) \cr
    & 4.61 $\pm$ 0.16 (2) \cr
    & 4.34 $\pm$ 0.06 (3)\cr
 metallicity ([Fe/H]) & 0.19 $\pm$ 0.03 (1)\cr
    & 0.26 $\pm$ 0.06 (2) \cr
    & 0.11 $\pm$ 0.03 (3) \cr
\hline
\end{tabular}
\end{flushleft}
References : (1) Gonzalez et al. \cite{gonzalez01}, (2) Santos et al. \cite{santos04}, (3) Fischer \& Valenti \cite{fischer05}.
\end{table*}

It is well known from the so-called ``asymptotic theory of stellar oscillations" (Tassoul \cite{tassoul80}) that modes with the same $\ell$ values and n numbers increasing by one are nearly equally spaced in frequencies. This spacing is called ``large separation". For $\iota$ Hor, the large separation deduced from the observations is 120 $\mu$Hz.  In Fig. 3, the identified modes are displayed in an ``echelle diagram", which is obtained by plotting their frequencies in ordinates, and the same frequencies modulo the large separation in abscissas. Should the large separation really be constant, the symbols representing the modes would lie on four vertical lines, for different $\ell$ values: from left to right, $\ell$~=~2,~0, and 3,~1. In real stars, the large separation varies slowly, so that the lines are curved and not exactly parallel.

\begin{figure}
\includegraphics[width=9cm]{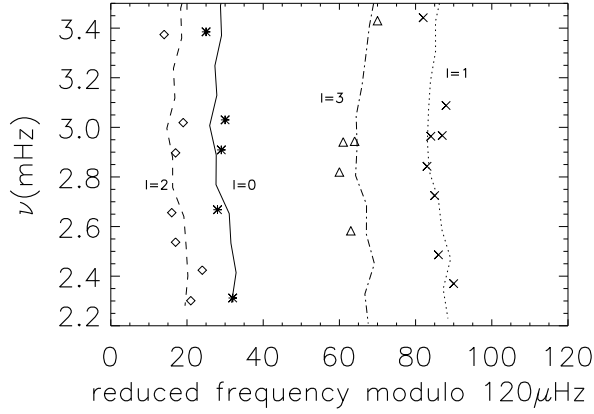}
\caption{ Observed and computed echelle diagrams. The echelle diagram displays the frequencies of the oscillating modes on ordinates, and the same frequencies modulo the large separation on abscissas. The symbols correspond to the identified modes (aliases have been taken off), and the continuous lines correspond to one of our best models (the third one of those presented in Table 2).}
\label{fig3}
\end{figure}

The distance between lines corresponding to $\ell$ and $\ell~+~2$ are called the ``small separations". In Fig.~3, it can be seen that the small separations between the $\ell$~=~2 and $\ell$~=~0 lines are of order 11 $\mu$Hz, which is close to the ``alias of the day", i.e. the frequency corresponding to 24 hours. As the night observations are separated by such a period, secondary peaks called ``aliases", are observed on each side of ``real" peaks. In particular, the peaks obtained here for $\ell$ = 0 include some aliasing effect from the $\ell$ =2 modes and the reverse. However, our observations show that the amplitudes of the $\ell$ = 0 and $\ell$ = 2 peaks have the same magnitude, while the day aliases have amplitudes that are only 3/4 of the signal: the identified peaks definitely contain a real signal. A complete analysis of the seismic modes for this star, as well as a general discussion of the models, will be given in a forthcoming paper. In Fig. 3, the aliases have not been plotted for the sake of clarity.

\section{Models and seismic analysis}

Stellar models were computed with the Toulouse-Geneva Evolutionary Code (TGEC) (Hui Bon Hoa \cite{hui07}). Various metallicities, helium values, effective temperatures, and gravities were tested. The oscillation frequencies for these models were computed using the adiabatic code PULSE (Brassard et al. \cite{brassard92}) and were compared to the observational data. More precisely, several evolutionary tracks, corresponding to different masses, were computed for each metallicity, as derived by the observers (Table 1). However, as already pointed out in Laymand \& Vauclair (\cite{laymand07}), no model can be computed for the Santos et al. (\cite{santos04}) values, as the corresponding error box ($\delta$ T$_{eff}$~-~$\delta$ log g) lies below the zero age main sequence. We computed models with metallicities [Fe/H] = 0.19 and 0.11, and for better precision in our results, we added evolutionary tracks for another intermediate metallicity: [Fe/H] = 0.14.

For each metallicity, computations were done for three different helium values. In the first case, the helium mass fraction Y was chosen proportional to the heavy elements mass fraction Z, as derived for the chemical evolution of galaxies (Isotov \& Thuan \cite{isotov04}). In the second case, Y was given an intermediate value, as given for the Hyades by Pinsonneault et al. \cite {pinsonneault04} (Y=0.271); and in the third case, Y was as low as found by Lebreton et al. (\cite{lebreton01}) for the Hyades (Y=0.255).

For each track, the only model fitting the observed large separation was extracted. For each couple ([Fe/H], Y), only one of these models also fitted the small separations. In other words, only one model can fit a given metallicity, a given helium abundance, and the observed seismic frequencies all together. These ``best models" are displayed in Fig. 4. Taking the observed uncertainties on the frequency determinations into account (1 $\mu$Hz), the uncertainties on the effective temperatures and gravities of the models that can fit the seismic observations may be evaluated as: $\delta$ T$_{eff} \simeq 10$K and $\delta$ log g $\simeq$ 0.01, which is the size of the symbols used in Fig.4. The gravities of these models vary between log g = 4.398 and 4.403, inside the evaluated uncertainty. Their ages are between 600 and 700 Myr, and their masses between 1.21 and 1.26 M$_\odot$. Their luminosities lie between log~(L/L$_\odot$)~=~.237 and 0.265; they decrease for decreasing helium and increase for decreasing metallicity.

\begin{figure}
\includegraphics[width=9cm]{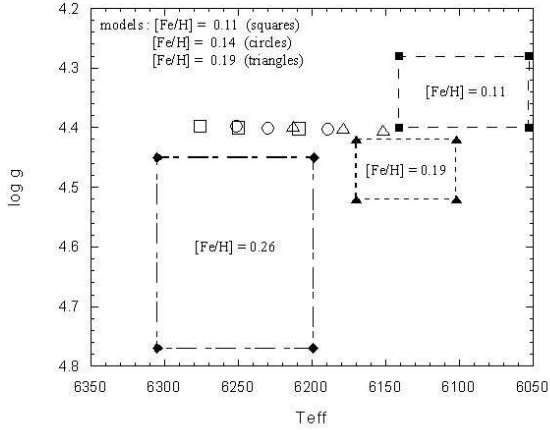}
\caption{ In the T$_{eff}$-log g plane, the three spectroscopic boxes presented in Table 1 are displayed: Santos et al. \cite{santos04} (box~[Fe/H]~=~0.26); Gonzalez et al. \cite{gonzalez01} (box~[Fe/H]~=~0.19); Fischer \& Valenti \cite{fischer05} (box~[Fe/H]~=~0.11). The open symbols correspond to the nine models which fit both the large and small separations. They correspond to three metallicities: [Fe/H]~=~0.11, 0.14, and 0.19 and three helium values: Y proportional to Z, Y~=~0.271, and Y~=~0.255. As discussed in the text, no model can be computed for the metallicity [Fe/H]~=~0.26, as the corresponding box lies below the ZAMS. Inside a metallicity group (same symbols), the model with the highest Y value is on the left, the one with the lowest Y value on the right.}
\label{fig3}
\end{figure}

As is clearly visible in Fig. 4, for a given metallicity the models are cooler for smaller helium abundances, and for a given helium abundance, the models are cooler for a higher metallicity. The parameters for the three coolest models are given in Table~2. Among all the computed models, they are the ones for which the effective temperatures and gravities are the closest to the spectroscopically determined ones, according to their metallicities. While all the models have slightly higher luminosities than the value derived from the Hipparcos data (Table~1), the three coolest models are those with the lowest luminosities, the closest to the observational upper limit.

\begin{table*}
\caption{iota Horologii: parameters of the three best models (columns 2, 3, and 4), compared to the Hyades values (column 5, reference: Lebreton et al. \cite{lebreton01})}
\label{tab2}
\begin{flushleft}
\begin{tabular}{ccccc} \hline
\hline
 [Fe/H] & 0.19 & 0.19 & 0.14 & 0.14 $\pm$ 0.05   \cr 
 Y & 0.271 & 0.255 & 0.255 & 0.255 $\pm$ 0.013 \cr
 age (Myr) & 620 & 627 & 627 & 625 $\pm$ 25 \cr
 mass (M$_\odot$) & 1.24 & 1.26 & 1.25 & \cr
 Teff (K) & 6179 & 6136 & 6189 & \cr
 Log g & 4.40 & 4.40 & 4.40 & \cr
 Log L/L$_\odot$ & 0.245 & 0.237 & 0.250 & \cr
\hline
\end{tabular}
\end{flushleft}
\end{table*}

These results lead to the following conclusions for the star iota Horologii:
\begin{itemize}

\item (Fe/H) is between 0.14 and 0.19
\item Y is small, $\simeq 0.255 \pm 0.015$
\item the age of the star is $625 \pm 5$ Myr
\item the gravity is $4.40 \pm 0.01$
\item the mass is $1.25 \pm 0.01$ M$_\odot$

\end{itemize}

The values obtained for the metallicity, helium abundance and age of this star are those characteristic of the Hyades cluster.

\section{Discussion}

This work shows how powerful asteroseismology can be in deriving the characteristic parameters of a star with the help of spectroscopic analysis, but finally obtaining much more precise results than with spectroscopy alone. In the case of iota Horologii, alias HD 17051, we have derived precise metallicity, helium abundance, age, gravity, and mass. We have shown that the values of [Fe/H], Y, and age are the same as for the Hyades cluster.

That the star iota Horologii, situated in the southern hemisphere, not only belongs to the Hyades stream, but also has the same age, metallicity, and helium abundance as the stars of the Hyades cluster itself, cannot be due to chance. We claim that this star belongs to the minority of the stream stars that have actually been evaporated from the primordial cluster.

An important consequence of $\iota$ Hor being a Hyades-evaporated star concerns the origin of its overmetallicity. 
As pointed out by several authors (Santos et al. \cite{santos03} and \cite{santos05}, Gonzalez \cite{gonzalez03}, Fischer \& Valenti \cite{fischer05}), 
exoplanet-host stars are, on average, overmetallic with respect to the Sun by a factor two. 
The origin of this overmetallicity has not yet been completely settled. It is probable that the clouds out of which the stars were born were already overmetallic, but it may also happen that the stars accrete many planets and planetesimals during their early phases, thereby increasing their surface metallicity. Note that, in contrast to the argument raised by Santos et al. \cite{santos03}, that the observed overmetallicity presents no correlation with spectral type does not contradict accretion, as shown by Vauclair (\cite{vauclair04}): due to thermohaline convection, the resulting metallicity would be related to the $\mu$-gradient inside the star, not to the initial conditions.

If the overmetallicity is primordial, it extends throughout the star, from the surface to the centre. 
If it is due to accretion, the star is overmetallic only in its outer layers. 
Asteroseismology should be able to distinguish between these two cases (Bazot \& Vauclair \cite{bazot04}), 
but it was not yet possible to obtain unambiguous results, even for the well-studied star $\mu$ Arae, 
which was observed with HARPS with a high precision (Bouchy et al. \cite{bouchy05}, Bazot et al. \cite{bazot05}).

In the case of $\iota$ Hor, the question may be approached in a different way. 
Asteroseismology leads to the conclusion that this star was evaporated from the primordial Hyades cluster, sharing the same age, helium abundance and overmetallicity. 
This is one proof that the origin of this overmetallicity is primordial, from the original cloud. This result has important consequences for the formation and subsequent evolution of galactic clusters and the theories of exoplanets formation and migration.

\begin{acknowledgements}

We thank the referee for very useful comments and important suggestions. 

\end{acknowledgements}

\end{document}